\documentclass{jpp}
\usepackage{graphicx}
\usepackage{epstopdf}
\usepackage[utf8]{inputenc}
\usepackage{amsmath}
\usepackage{siunitx}
\usepackage{wasysym}
 
\title{Effect of scrape-off losses and  current flattening on runaway electron dynamics in ARC}
\author{William Weeks\aff{1},   István Pusztai\aff{1}, Ida Ekmark\aff{1},   Cesar Clauser\aff{2}, Ryan Sweeney\aff{3}, R.~Alex Tinguely\aff{2}
  \and  Tünde Fülöp\aff{1}  }

\affiliation{
\aff{1}Department of  Physics and Astronomy, Chalmers University of Technology, Gothenburg,  Sweden
\aff{2} Plasma Science and Fusion Center, MIT
\aff{3} Commonwealth Fusion Systems}

\begin{document}

\maketitle

\begin{abstract}
Disruptions in high-current tokamaks, such as ARC, may result in megaampere-scale runaway electron currents.   In this work, we evaluate mitigation strategies via massive material injection, accounting for the combined impact of scrape-off losses due to vertical displacement events and current profile relaxation during the thermal quench. We find that scrape-off losses alone yield only modest reductions in runaway electron currents, but combining them with current flattening creates a beneficial synergistic effect. The latter broadens the current channel outward, shifting efficient runaway electron generation closer to the plasma edge where the dynamically contracting boundary can effectively scrape off the electron beams. This interplay significantly expands the operational window for mitigation toward lower assimilated densities, theoretically reducing runaway currents to manageable levels at experimentally feasible assimilated quantities. 
\end{abstract}


\section{Introduction}
\label{sec:intro}
One of the challenges in designing a viable tokamak fusion power plant
is the mitigation of plasma disruptions; events where thermal and
magnetic confinement are rapidly lost \citep{Hender2007,Sweeney2020}. Disruptions pose severe risks
to the structural integrity of a reactor via three primary mechanisms:
intense, localized heat loads on plasma-facing components during the
thermal quench (TQ); large electromagnetic forces from eddy and halo
currents during the current quench (CQ); and the generation of a
relativistic beam of runaway electrons (REs) capable of causing damage upon wall impact \citep{hollmann_lims}.

To mitigate these hazards, baseline disruption mitigation systems (DMS)
for next-step devices rely on massive material injection (MMI) to rapidly
radiate away thermal energy and control the subsequent current decay.
Such mitigation strategies are being extensively studied on
present-day tokamaks \citep{Commaux2011,Bakhtiari2011,Lehnen2015a,REUX2015143,AsdexUpgradeTeam2015,Pautasso2017, Eidietis2017,Shiraki2016,Commaux2016,Herfindal2019,
Sheikh2021,Sheikh2024,Jachmich2022}. However, scaling them to future
reactor-scale devices involves physics that
cannot be fully replicated in existing machines. Consequently,
predictive numerical modelling is indispensable. Full
three-dimensional (3D) magnetohydrodynamic (MHD) codes offer
comprehensive insights into these events but are computationally
infeasible for wide-ranging parameter scans. This has motivated the
widespread use of reduced-dimensionality tools, such as the 1D
fluid-kinetic framework DREAM \citep{dreampaper}.

Historically, many 1D disruption studies have operated under the
conservative assumption of a fixed magnetic flux-surface geometry,
neglecting vertical plasma motion and subsequent scrape-off
losses. This omission typically leads to an overestimation of the
final RE current relative to experiment. Because open field lines cannot sustain relativistic
beams, vertical displacement events (VDEs) scrape off the outer layers
of the plasma, rapidly exhausting REs to the wall.

Axisymmetric MHD simulations have demonstrated that this vertical
motion restricts the variation of the poloidal magnetic flux inside
the remaining last closed flux surface (LCFS) \citep{Wang2025}. Given that
the avalanche enhancement factor depends exponentially on this flux
variation \citep{Rosenbluth_1997}, accounting for VDE-driven scrape-off is
important for accurate RE prediction. Recent investigations indicate
that incorporating a simplified, flux-conservative scrape-off model
can radically alter predictions, making  RE avoidance
theoretically possible in certain regimes
\citep{Vallhagen2025,Votta_2026}.

Recent exploratory modelling utilizing multi-objective Bayesian
optimization showed persistent megaampere-scale runaway beams in ARC V3A \citep{ARC2026}. Throughout this paper, ARC refers to the V3A design. However,
these simulations omitted scrape-off effects and current
relaxation induced by stochastization of the magnetic field. The latter
phenomenon can be captured via a hyperresistive transport term in
Faraday's law \citep{Boozer1986,Boozer_2018}.

To assess mitigation performance, we use three reference values as general
metrics of success: a maximum runaway electron current
$I_{\rm RE} < 150\,\rm kA$, a conducted fraction of the pre-disruption thermal
energy $\eta_{\rm cond} < 10\%$, and a current quench time in the range
$11.3\,{\rm ms} < \tau_{\rm CQ} < 142\,{\rm ms}$. These are not ARC design
requirements or targets. The $\tau_{\rm CQ}$ range is
motivated by the electromagnetic-load considerations in \citet{ARC2026}.
The $\eta_{\rm cond}$ threshold is our own choice; it serves as a simplified
proxy for the actual ARC requirement of avoiding melting of plasma-facing
components, which has not been expressed as a limit on the conducted or
radiated energy fraction. The $I_{\rm RE}$ threshold is chosen
by the authors following earlier mitigation studies such as
\citet{ARC2026}.

In this paper, we utilize the DREAM code \citep{dreampaper}, including a hyperresistive
current-relaxation model and a flux-conservative scrape-off framework \citep{Vallhagen2025},
to re-evaluate mitigated disruption scenarios for ARC. Our findings reveal that while scrape-off alone yields modest reductions, its combination with current flattening expands the operational window for successful disruption mitigation towards lower assimilated densities, potentially suppressing runaways to manageable levels. However, within the heat transport model used here, which likely overestimates
conducted losses, satisfying the reference value for conducted heat losses
would require larger assimilated quantities than are optimal for runaway
suppression. 

The remainder of this paper is organized as follows. In Section 2, we
outline the numerical framework, describing the VDE scrape-off model
and the hyperresistive current relaxation in DREAM. Section 3 presents
the results of the simulations for
ARC, contrasting cases with and without current relaxation and
scrape-off effects.
We summarize our conclusions in Section \ref{sec:concl}.

\section{Disruption modelling}
\label{sec:theory}

To evaluate the combined effects of vertical plasma motion and current-profile relaxation on RE dynamics during disruptions in ARC, we employ
the 1D fluid-kinetic framework DREAM \citep{dreampaper}. The
disruption scenario modelled here mirrors the baseline activated ARC V3A
simulation setup described by \citet{ARC2026}, modified to account for a current-density relaxation during the TQ and scrape-off effects.

Following the baseline configuration for an activated ARC scenario, we
assume an initial plasma current $I_{\rm p} = 12\,\mathrm{MA}$ and an
on-axis toroidal magnetic field $B_0 =11.3\,\mathrm{T}$
\citep{ARC2026}. The resistive wall time is set to $\tau_{\rm w} = 250\,{\rm ms}$.  
Disruption mitigation is initiated via MMI, assuming instantaneous and radially uniform deposition of mixed deuterium (D) and neon (Ne) impurities. The prescribed densities represent assimilated material and are input parameters; the injection and assimilation processes are not explicitly resolved. Throughout this work, we distinguish these assimilated densities from the quantities supplied by the injection system, of which only a fraction may be assimilated.

The thermal quench is triggered by prescribing an elevated
Rechester-Rosenbluth-type diffusive transport of heat,
superthermal electrons and runaway electrons.
The radial heat transport of electrons is calculated assuming a parallel heat flux scale of $q_\| \sim n_e v_{te} T_e$ along the perturbed field lines, with $v_{te}=\sqrt{2T_e/m_e}$ the electron thermal speed, $n_{\rm e}$ and $T_{\rm e}$ the electron density and temperature, and $m_e$ the electron mass. 
We note that kinetic simulations of the thermal collapse of a nearly collisionless plasma along field lines intersecting a cooling region find that, as a consequence of ambipolarity, the bulk of the electron and ion thermal energy is transported at speeds set by the ion sound speed rather than the electron thermal speed \citep{Zhang2023}. A small fraction of fast electrons leaves promptly and sets up a potential that confines the rest. The electron-thermal-speed scaling adopted here is common in reduced disruption models, but it is therefore likely to overestimate conducted heat losses. The implications for the conducted heat fraction are discussed in Section 3.1.

The diffusion coefficient corresponds to a spatially and temporally constant, normalized magnetic perturbation of $\delta B/B= 0.3 \%$. It is applied until the temperature falls below $20\,\rm eV$ on radial average, which provides our definition of the end of the thermal quench (this time instance is denoted by $t_{\rm TQ}$). The conducted heat loss $\eta_{\rm cond}$ is defined as the fraction of the thermal energy lost through heat transport during the thermal quench. The loss is evaluated up to the time when the post-TQ temperature is reached. During the TQ, a diffusive runaway-electron transport corresponding to the same value of $\delta B/B$ is also applied. 

The evolution of the plasma species is resolved by tracking the
ionization and recombination dynamics of the background ions,
the impurities, and the bulk electron populations. To accurately capture
the hot-tail seed generation mechanism during the TQ, the electrons
are divided into three populations: the bulk thermal
background (also referred to as the ``cold'' population) and the runaway electrons are modelled using appropriate fluid
descriptions.  The population at intermediate momenta, that includes typical values of the critical momentum during the TQ, referred to as the ``hot'' population, is modelled using a pitch-angle-averaged kinetic framework, as in \cite{ekmark2024}, which provides an accurate model for seed generation, not only with respect to the hot-tail and Dreicer processes, but also the activated seed generation mechanisms. This
approach balances computational tractability with the high fidelity
required to capture early-stage seed generation. We do not expect a fully kinetic treatment of the runaway population
to significantly change the runaway current evolution predicted here.
Such a treatment would provide information about the runaway electron
energy distribution, which is relevant for assessing wall loads but
is beyond the scope of this work. Moreover, the fully kinetic model
in DREAM uses a collision operator expanded about a single Maxwellian
background. It is therefore less suited to describing the initial
temperature equilibration between the cold injected material and
the pre-existing plasma, which our treatment accounts for.

\subsection{VDE scrape-off model}
MHD simulations across various reactor designs, including M3D-C1
simulations for ARC \citep{Clauser_2025}, demonstrate that during the
runaway-generation phase, the poloidal magnetic flux\footnote{The poloidal flux $\psi$ is defined according to the convention used by \citet{Boozer_2018}, between the magnetic axis and the innermost point of the flux surface.} $\psi$  at the
last closed flux surface (LCFS) remains approximately constant, i.e.,
$\psi(r_{\mathrm{LCFS}},t) \approx \psi(a,t=0)$, where $a$ is the minor radius of the plasma. 
This
behaviour stems from the high conductivity of the first wall, especially if 
the distance between the wall and the plasma is small, as in ARC.
Any flux changes induce strong currents in the wall that act to cancel
the flux variation. To avoid the substantial computational expense of
RE evolution coupled to 3D MHD simulations while retaining the physics
of VDEs, we use a reduced scrape-off model in DREAM, described by
\citet{Vallhagen2025}.

At each computational time step, the instantaneous radius of the last
closed flux surface, $r_{\mathrm{LCFS}}(t)$, is determined by
numerically solving the equation:
\begin{equation}\label{eq:lcfs_flux}
\psi(r_{\mathrm{LCFS}}, t) = \psi(a,0),
\end{equation}
where the right hand side represents $\psi$ at the prescribed radius of the last closed flux surface at $t=0$.  
The region outside the boundary $r > r_{\mathrm{LCFS}}$ represents
open magnetic field lines intersecting the first wall. REs entering
this zone are rapidly lost, which is enforced numerically by a loss
term that is finite only for $r > r_{\mathrm{LCFS}}$:
\begin{equation}\label{eq:REloss_term}
\frac{\partial n_{\mathrm{RE}}}{\partial t} = - \frac{n_{\mathrm{RE}}}{\tau_{\mathrm{loss}}}\Theta (r-r_{\mathrm{LCFS}}),
\end{equation}
where $\Theta$ is the Heaviside step function, and the exact value of the loss timescale $\tau_{\mathrm{loss}}$ is unimportant as long as it is significantly (in this case several orders of magnitude) shorter than any other time scale of interest; we set it to $\tau_{\mathrm{loss}}=5\Delta t$, where $\Delta t$ is the longest allowed simulation time step. 
We also track the cumulative contribution of scrape-off losses, denoted by $I_{\text{RE,lost}}$, by accumulating the current contributions removed by this loss term over time, while a finite confined region remains. This quantity expresses the accumulated runaway losses in current units; it does not represent a current that continues to flow after the electrons have been lost. In contrast, in simulations with scrape-off enabled (denoted by
$\text{+SO}$), $I_{\text{RE,+SO}}$ denotes the instantaneous confined runaway current.

A similar loss term is also included on the electron thermal energy density as
\begin{equation}\label{eq:Wloss_term}
\frac{\partial W_{\mathrm{e}}}{\partial t} = - \frac43\frac{\gamma_{\rm sh}n_{\rm e}c_{\rm s}T_{\rm e}}{L_\parallel}\Theta (r-r_{\mathrm{LCFS}}),
\end{equation}
where $\gamma_{\rm sh}=8$ is the heat transmission coefficient, $c_{\rm s}$ the sound speed and $L_\parallel$ the length of the flux tube along the magnetic field~\citep{Votta_2026}. 
This term approximates parallel heat losses along open field lines
in the halo region using a prescribed sheath heat transmission
coefficient. The factor $n_{\rm e}c_{\rm s}$ represents the particle
flux scale associated with the ion saturation current,
$j_{\rm sat}\simeq e n_{\rm e}c_{\rm s}$. The halo current and its
effect on sheath heat transmission are not explicitly resolved
in this reduced model. For sufficiently smooth radial dynamics and converged current evolution we use $N_r=80$ radial grid points. 

\subsection{Hyperresistive current profile relaxation}
During the TQ, MHD instabilities stochastize the magnetic field, which
drives a rapid relaxation (i.e., flattening) of the current density profile, manifested
as a distinct spike in the total plasma current $I_{\rm p}$ and a sharp reduction in the internal inductance $l_{\rm i}$. We model the spatial redistribution of the current density by a hyperdiffusive term in Faraday's law, which is constructed to conserve the total magnetic helicity $\propto \int \psi(\psi_{\rm t},t)d\psi_{\rm t}$ \citep{Boozer_2018,PusztaiHyperres}, where $\psi_{\rm t}$ denotes the toroidal flux. The evolution of the poloidal flux is governed by
 \begin{equation}\label{eq:faraday_hyper}
\frac{\partial \psi}{\partial t} = - {V_\text{loop}} +
\frac{\partial}{\partial \psi_{\text{t}}} \bigg(\psi_{\text{t}}
\Lambda \frac{\partial}{\partial \psi_{\text{t}}}
\frac{j_{\|}}{B}\bigg),\end{equation} where $j_\parallel$ is
 the total parallel plasma current density, and $\Lambda(r,t)$ represents the anomalous hyperdiffusion coefficient, commonly termed the
 \emph{hyperresistivity}, which is non-zero in regions where the magnetic field lines are stochastic. Furthermore, $V_\text{loop}=\eta_\|(j_\Omega/B)2 \pi\langle B^2\rangle/\langle\nabla \varphi\cdot \mathbf{B}\rangle$ represents poloidal flux evolution due to ohmic dissipation, with the parallel ohmic current $j_\Omega$ and resistivity $\eta_\|$, and $\langle\cdot\rangle$ denoting flux-surface average. The parallel electric field is calculated solely from this term (hence the notation $V_{\rm loop}$) as $E_\|=\eta_\|(j_\Omega/B)\sqrt{\langle B^2\rangle}$, since the volume-integrated parallel electric field vanishes in stochastic regions. Consequently, electrons that sample this volume along stochastic field lines experience no net acceleration, as pointed out by \citet{Boozer_2017}.   
 The hyperresistive term is implemented with zero radial helicity flux at the outer boundary of the fixed computational domain, so that this term conserves the domain-integrated magnetic helicity. This condition is distinct from the resistive-wall boundary condition governing poloidal flux evolution and is not imposed at the moving LCFS used in the scrape-off model.

The magnitude of the hyperresistivity can be estimated following \citet{Boozer_2018} as
\begin{equation}
\Lambda\approx\frac{1}{144}\frac{\mu_0}{4\pi}\frac{V_{\rm A} \psi_{\rm t}^2}{N_{\rm t}},
\label{eq}
\end{equation}
where $\mu_0$ is the vacuum permeability, $V_{\rm A}=B/\sqrt{\mu_0\rho}$ is a representative Alfvén speed, $\rho$ is the mass density, and $N_{\rm t}$ denotes the number of toroidal transits required for a stochastic field line to sample the entire radial domain. The otherwise unknown parameter $N_{\rm t}$ can be related to the magnetic perturbation strength $\delta B/B$ used to model the transport of heat and energetic electrons \citep{Votta_2026}, yielding
\begin{equation}
\Lambda \approx \frac{\mu_0 \pi q R_0^2 V_{\rm A} \psi_{\rm t}^2}{288 a^2}
\left(\frac{\delta B}{B}\right)^2,
\label{vottalambda}
\end{equation}
where $a$ and $R_0$ are the minor and major radii of the plasma, respectively, and $q$ is the safety factor. To obtain this relation, we identify the characteristic parallel distance travelled by a runaway electron before diffusing out of the plasma with $L=2\pi R_0N_{\rm t}$. This distance can also be estimated from the characteristic diffusive loss time, $\tau_{\rm RE}\sim a^2/D_{\rm RE}$, as $L\sim c\tau_{\rm RE}$. Using the Rechester--Rosenbluth-type estimate $D_{\rm RE}\sim\pi qR_0c(\delta B/B)^2$ therefore gives
$
N_{\rm t}\sim a^2 /[2\pi^2 qR_0^2 \left(\delta B/B \right)^{2}],
$
which, when inserted into Eq.~\eqref{eq}, yields Eq.~\eqref{vottalambda}. Based on the estimate Eq.~(\ref{vottalambda}), typical values of $\Lambda_0$ are of order unity in units of $\rm Wb^3/C$. At this level of hyperresistivity, the effects of current relaxation are effectively saturated: varying $\Lambda$ by factors of order unity does not appreciably change the results.  Furthermore, the radial variations of $\Lambda$ considered in Appendix~\ref{sec:AppendixSpatialVariation} have only a small effect on the resulting runaway currents. Finally, we note that $\Lambda$ is set to zero outside the thermal quench in our simulations.

This formulation allows a synergistic interplay:
hyperresistive current relaxation flattens $j_\parallel$, broadening the
current channel outward, which in turn shifts efficient runaway generation towards the edge, where it is subsequently scraped off by the dynamically contracting boundary.

\section{Runaway generation in the presence of massive material injection}
\label{sec:results}

\subsection{Mitigation efficacy vs.~assimilated material quantities}

\begin{figure}
  \begin{center}
  \includegraphics[width=7.92cm]{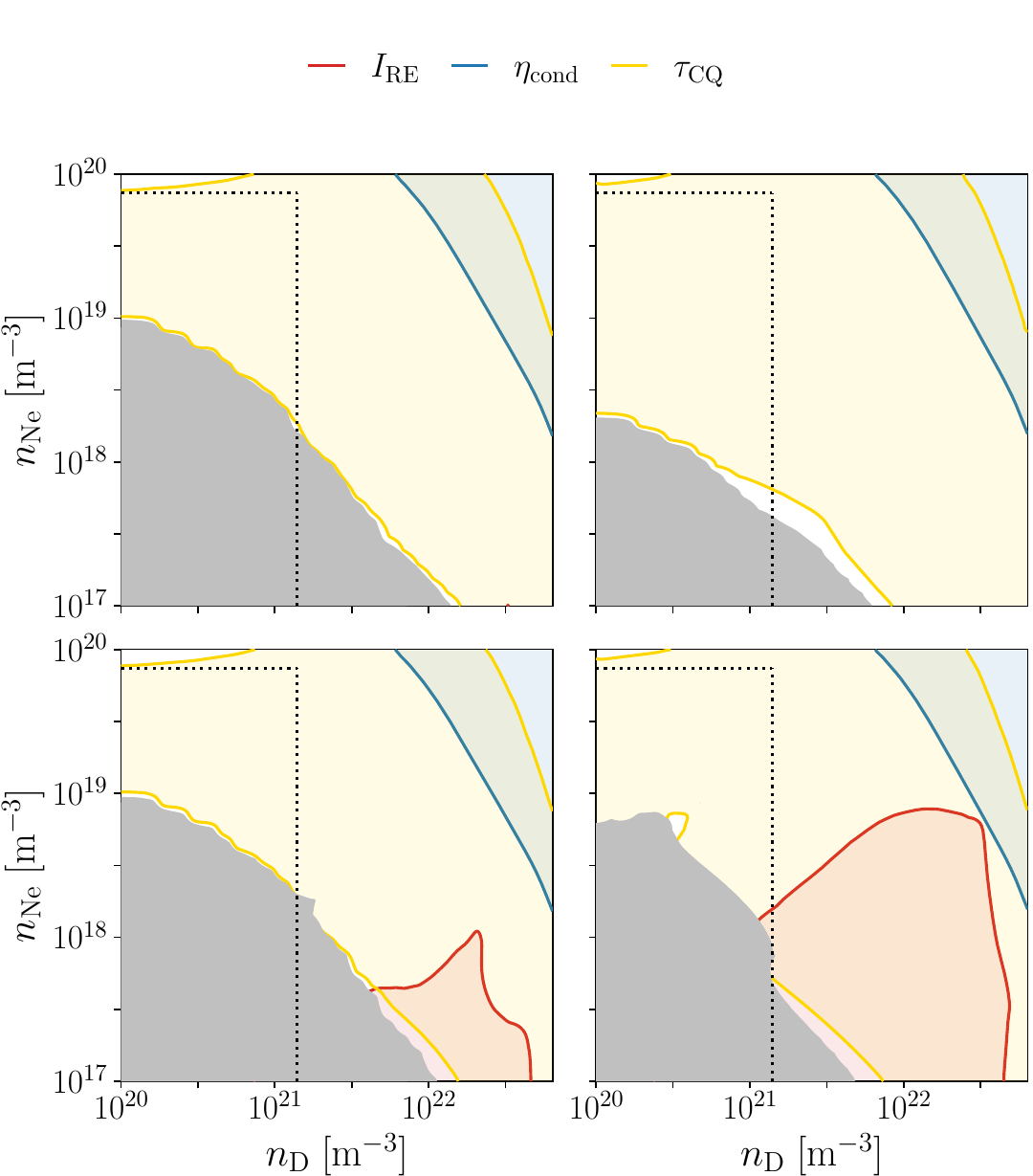}
  \put(-195,131){(a)}
  \put(-195,141){\tiny -SO,-CR}
  \put(-92, 131){(b)}
  \put(-92, 141){\tiny -SO,+CR}
  \put(-195,27){(c)}
  \put(-195,37){\tiny +SO,-CR}
  \put(-92,27){(d)}
  \put(-92,37){\tiny +SO,+CR}
  \end{center}
  \caption{Regions satisfying the reference values (colour shaded areas) for the conducted heat fraction  ($\eta_{\text{cond}}<10 \%$, blue),
 current quench time ($11.3\,{\rm ms}<\tau_{\text{CQ}}< 142\,{\rm ms}$, yellow), and maximum runaway electron current  ($I_{\text{RE}}<150\;\rm kA$, red), in the parameter space of assimilated densities. (a) and (b) are without, (c) and (d) are with scrape-off (-SO and +SO, respectively).
 (a) and (c) are without, (b) and (d) are with current relaxation (-CR and +CR). 
 The gray region indicates assimilated densities resulting in an incomplete TQ and CQ. The plotted parameter domain is covered by $32\times 32$ simulations, equidistant on the logarithmic scales. Estimated upper limits of $n_{\rm D}$ and $n_{\rm Ne}$ at planned MMI capabilities and an assumed $10\%$ assimilation rate are indicated with the dotted rectangular boundary.}
\label{fig:contours}
\end{figure}

We start by evaluating disruption mitigation efficacy across a parameter space
of assimilated deuterium and neon densities, using the reference values for
$I_{\rm RE}$, $\eta_{\rm cond}$ and $\tau_{\rm CQ}$ introduced in
Section~\ref{sec:intro}. The current quench time is evaluated as described by
\citet{ekmark2024}.

Parameter regions satisfying these reference values are highlighted and bounded by coloured contours in Fig.~\ref{fig:contours}; red corresponding to $I_{\rm RE}$, blue to $\eta_{\rm cond}$, and yellow to $\tau_{\rm CQ}$. The gray region in the lower-left corner,  with a boundary close to the $\tau_{\text{CQ}} = 142\,{\rm ms}$ limit, indicates an incomplete TQ.  An incomplete TQ occurs when a substantial fraction of the plasma ohmically reheats after the elevated transport window, sustaining a long-lived ohmic current. Thus, we use the term \emph{incomplete TQ} henceforth to denote cases in which ohmic reheating also prevents completion of the CQ.
The practical criterion we employ for incomplete TQ is $\tau_{\rm CQ}>500\,{\rm ms}$ or a remnant ohmic current at the end of the simulation (that is slightly longer than the upper $\tau_{\rm CQ}$ limit) higher than $\SI{1}{MA}$. The figure panels compare scenarios without (-SO, upper panels) and with (+SO, lower panels) scrape-off effects, alongside cases without (-CR, left panels) and with (+CR, right panels) hyperresistive current density relaxation.

The conducted heat  fraction reference value is met only in the high-D--high-Ne region of the parameter space, where radiated  losses are sufficiently large. The corresponding blue area in the upper right corner shows negligible sensitivity to either scrape-off or current density relaxation effects. Similarly, the $11.3\,{\rm ms}$ $\tau_{\rm CQ}$ limit (upper yellow curve in the figure), which emerges at high values of $n_{\rm D}$ or $n_{\rm Ne}$, is also fairly insensitive to these effects.
The $\eta_{\text{cond}}$ boundary should be interpreted with caution. If parallel energy transport is limited by the ion sound speed (Zhang et al. 2023), conducted losses would be reduced and radiation would dominate over a larger part of the parameter space. The $\eta_{\text{cond}}<10  \%$ region would then be expected to extend towards lower assimilated densities, possibly substantially. Sweeney et al. (2026) found that a sheath-limited heat-flux boundary condition substantially reduces $\eta_{\text{cond}}$ while leaving the runaway dynamics largely unchanged. We therefore expect the $I_{\text{RE}}$ results presented here to be considerably less sensitive to this model choice than $\eta_{\text{cond}}$.

On the other hand, the $142\,{\rm ms}$ $\tau_{\rm CQ}$ limit (lower yellow curve in the figure), which closely follows the incomplete TQ boundary, remains largely unaffected by scrape-off in the absence of current relaxation. However, when current relaxation is included, the region exceeding this limit shifts toward lower assimilated densities. 
The reason is that
re-heating predominantly takes place in the initially hotter, inner plasma regions where the current density is highest.  Flattening the current profile reduces the core ohmic current density, making the plasma less susceptible to reheating.  
Furthermore, incorporating hyperresistivity amplifies the effect of scrape-off, which is consistent with a radially broader reheated region resulting from the flatter current density profile.

The most significant changes occur in the parameter space where the runaway current remains below its reference value. Without scrape-off, this value is essentially unattainable   
 within the explored assimilated density ranges, except in the excluded incomplete TQ region.  Here, only the simulation at $n_{\rm D}=\SI{3.2e22}{m^{-3}}$ and $n_{\rm Ne}=\SI{1e17}{m^{-3}}$ has $I_{\text{RE}}<\SI{150}{kA}$ and only when current relaxation is neglected.  
With scrape-off there are successful mitigation zones  in the high-D--low-Ne quadrant of the parameter space.  In this regime, deuterium efficiently enhances collisional drag on REs, while the relatively slow plasma cooling and moderate post-TQ temperatures suppress hot-tail seeding and avalanches. 
This benefit diminishes at the absolute highest deuterium densities, where extremely low post-TQ temperatures trigger partial deuterium recombination, driving a rapid  increase in
$I_{\rm RE}$ as $n_{\rm D}$ increases further.    
Finally, incorporating current relaxation amplifies the effectiveness of scrape-off, significantly expanding the parameter space that successfully meets the  $I_{\rm RE}$ target.

However, even when accounting for both scrape-off and current flattening, no single parameter region simultaneously satisfies all three reference values. Figure \ref{fig:contours}d reveals a narrow operational point where the red and blue regions nearly intersect. Unfortunately, reaching this point requires a deuterium density that far exceeds the expected injection and assimilation capabilities of ARC's planned MMI systems.  
Assuming a  $10\%$ assimilation rate, the estimated upper operational limits are  $n_{\rm D}=1.4\times 10^{21}\,\rm{m}^{-3}$ and $n_{\rm Ne}=7.4\times 10^{19}\,\rm {m}^{-3}$, indicated by dotted lines in Fig.~\ref{fig:contours}.

\begin{figure}
  \begin{center}
  \includegraphics[width=7.92cm]{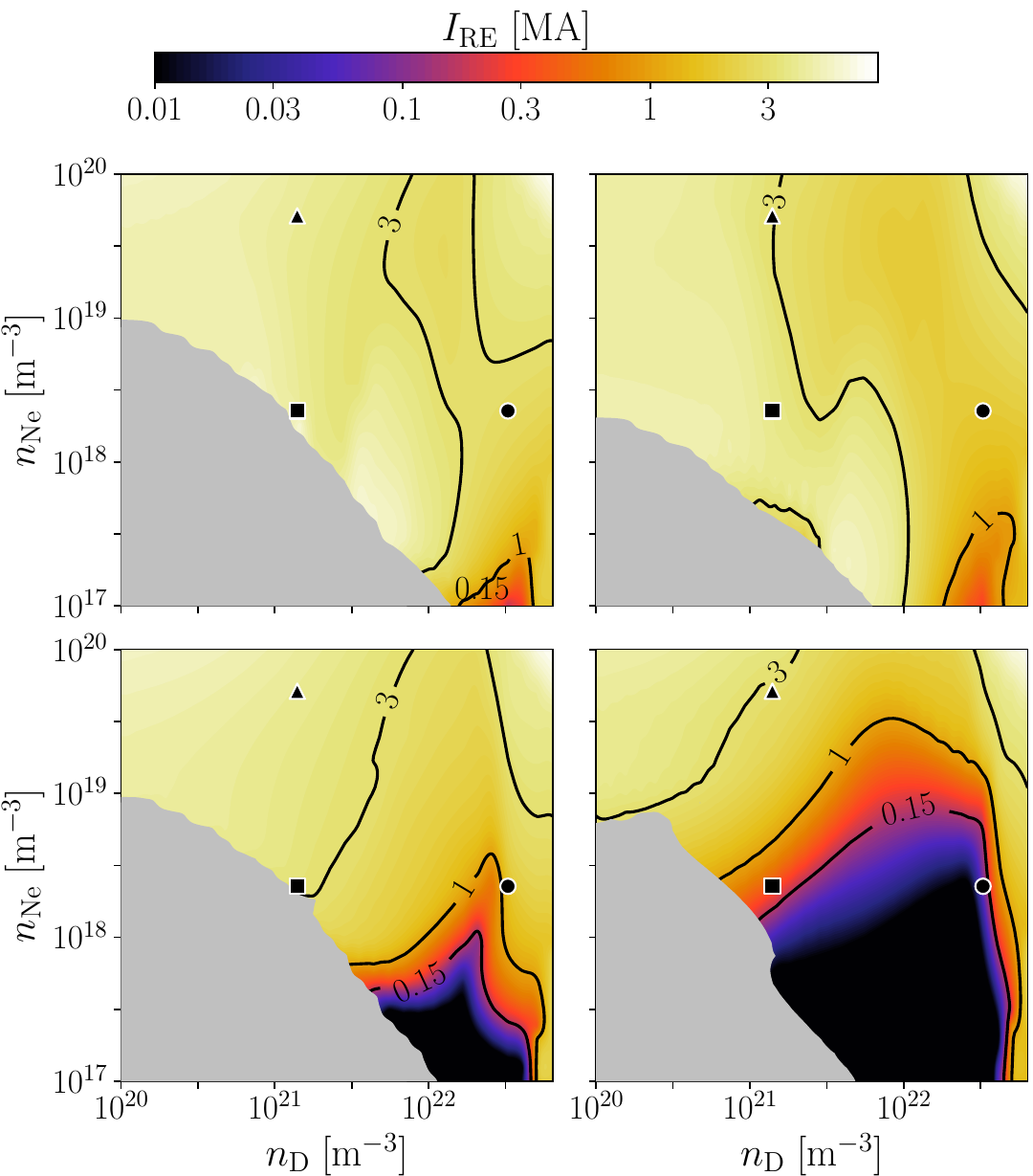}
  \put(-195,131){(a)}
  \put(-195,141){\tiny -SO,-CR}
  \put(-92, 131){(b)}
  \put(-92, 141){\tiny -SO,+CR}
  \put(-195,27){(c)}
  \put(-195,37){\tiny +SO,-CR}
  \put(-92,27){(d)}
  \put(-92,37){\tiny +SO,+CR}
  \end{center}
  \caption{Map of the predicted maximum runaway electron current as a function of the assimilated densities. (a) and (b) are without, (c) and (d) are with scrape-off (-SO and +SO, respectively). (a) and (c) are without, (b) and (d) are with current relaxation (-CR and +CR). 
 Contours denote specific runaway current levels
    ($0.15\,\text{MA}$, $1\,\text{MA}$, and $3\,\text{MA}$). The square (\scalebox{0.8}{$\blacksquare$}) indicates a case with $n_{\rm D} = 1.40 \times 10^{21} \, \si{m^{-3}}$ and $n_{\rm Ne} = 2.26 \times 10^{18} \, \si{m^{-3}}$. Compared to this, the triangle ($\blacktriangle$) case has an increased $n_{\rm Ne} = 5.13 \times 10^{19} \, \si{m^{-3}}$, and the circle (\raisebox{-0.3ex}{\scalebox{1.6}{$\bullet$}}) case has an increased $n_{\rm D} = 3.28 \times 10^{22} \, \si{m^{-3}}$. The gray region indicates assimilated densities resulting in an incomplete TQ and CQ. }
\label{fig:ire}
\end{figure}
\subsection{Runaway-electron current vs assimilated material}
Since $I_{\rm RE}$ demonstrates the highest sensitivity to scrape-off and current relaxation,  Fig.~\ref{fig:ire} examines this quantity in greater detail. In addition to the $0.15 \,\rm MA$ contour, which corresponds to the red boundary in Fig.~\ref{fig:contours}, $1\,\rm MA$ and $3\,\rm MA$  contours are included to complement the colour shading. These additions illustrate that for all scenarios, there is a region where $I_{\rm RE}$ drops below $1\,\rm MA$, though only for high deuterium and low neon densities when scrape-off is not considered. 

Without scrape-off (upper panels of Fig.~\ref{fig:ire}), current flattening is primarily beneficial above $n_{\rm Ne}=10^{18}\,\rm {m}^{-3}$, extending the $3\,\rm MA$ contour toward lower $n_{\rm D}$ values, 
but for $n_{\rm Ne}<\SI{3e17}{m^{-3}}$, $I_{\rm RE}$ is increased. The corresponding ``bubble'', a region where current flattening is ineffective in the absence of scrape-off, vanishes once scrape-off is enabled.  
When scrape-off losses are included, current flattening is universally beneficial within the explored parameter space, expanding the regions where $I_{\rm RE}$ lies below any given level. This is particularly clear for the three explicitly marked contours at $0.15$, $1$ and $3\,{\rm MA}$ in Fig.~\ref{fig:ire}c and d.

\subsection{Current evolution in representative cases}

In the following, we will analyze the runaway electron dynamics at three representative points within the assimilated density parameter space. The baseline case, denoted with the square marker in Fig.~\ref{fig:ire}, minimizes $I_{\rm RE}$, utilizing a practically achievable deuterium density ($n_{\rm D}\le 1.4\times 10^{21}\,\rm{m}^{-3}$). In this case the TQ is complete, regardless of scrape-off or current relaxation effects.
When both current relaxation and scrape-off are accounted for, {$I_{\rm RE} = \SI{349}{kA}$}; although this exceeds the 150 kA reference value, modern runaway electron experiments routinely manage currents of this magnitude. Additionally, a technically achievable scenario with higher assimilated neon content ($n_{\rm Ne} = 5.13 \times 10^{19} \, \si{m^{-3}}$,  triangle marker) is evaluated, despite offering no $I_{\rm RE}$ mitigation benefits.  Finally, a scenario with highly elevated deuterium content ($n_{\rm D} = 3.28 \times 10^{22} \, \si{m^{-3}}$, circle marker) serves as a theoretical reference point, even though it exceeds currently planned injection capabilities.  

\begin{figure}
    \begin{center}
        \includegraphics[width=11.88cm]{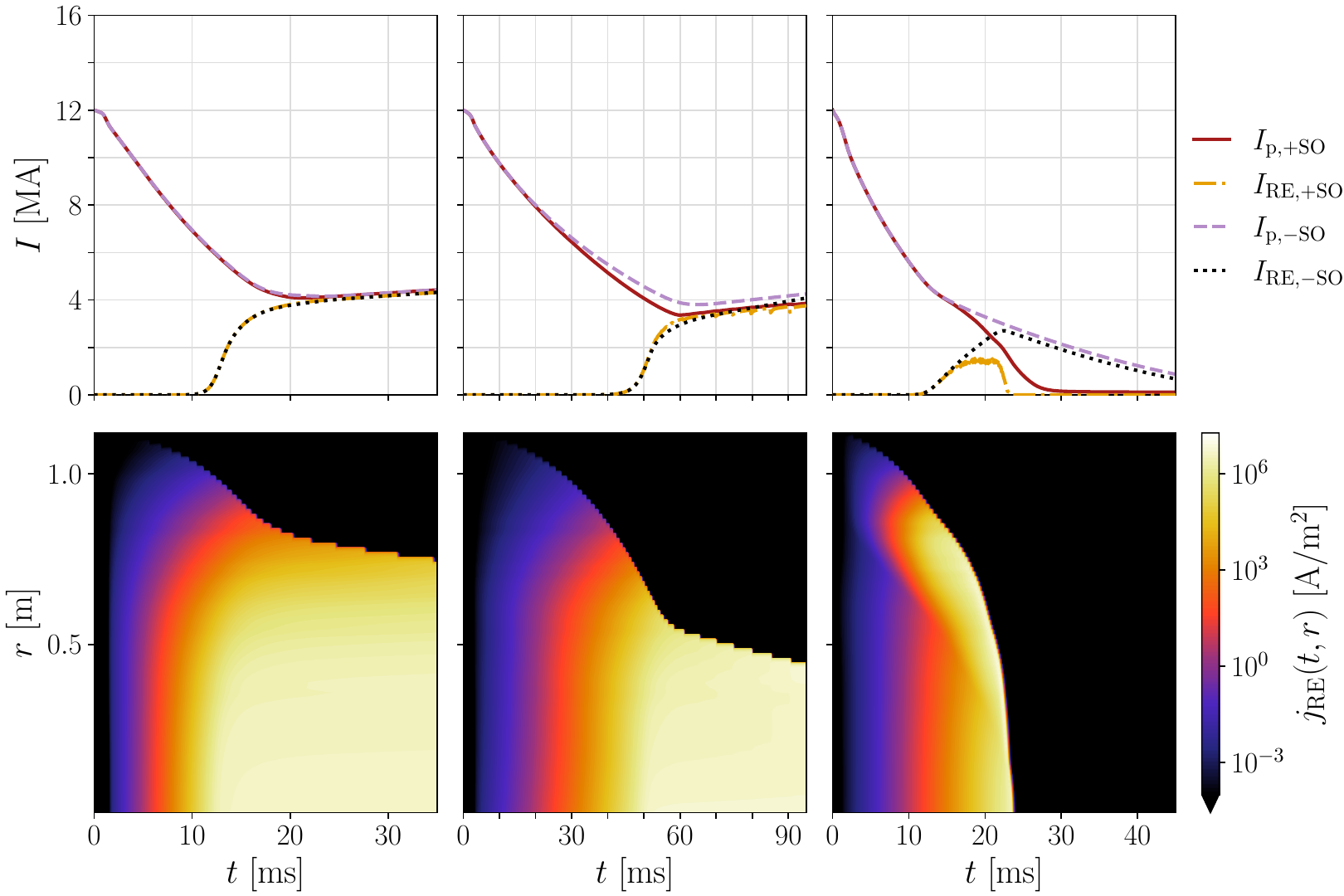}
        \put(-309,213){$\blacktriangle$} 
        \put(-215,213){\scalebox{0.8}{$\blacksquare$}}
        \put(-123,212){\scalebox{1.6}{$\bullet$}}
        \put(-243,212){(a)}
        \put(-150,212){(b)}
        \put(-56,212){(c)}
        \put(-243,106){\textcolor{white}{(d)}}
        \put(-150,106){\textcolor{white}{(e)}}
        \put(-56,106){\textcolor{white}{(f)}}
    \end{center}
    \caption{(a)-(c) Time evolutions of the total plasma currents and the runaway electron currents \textbf{without current density relaxation}, for the assimilated quantity combinations represented by the markers in Fig.~\ref{fig:ire}. Solid and dashed curves are total plasma current, dash-dotted and dotted are runaway currents. Dashed and dotted are without, solid and dash-dotted are with scrape-off.   (a) low-D--high-Ne ($\blacktriangle$); (b) low-D--low-Ne baseline (\scalebox{0.8}{$\blacksquare$}); and (c) high-D--low-Ne (\raisebox{-0.3ex}{\scalebox{1.6}{$\bullet$}}) scenario. 
    (d)-(f) Spatio-temporal evolution of the runaway current density with scrape-off, for the respective assimilated densities of panels (a)-(c). }
\label{fig:currents}    
    \end{figure}
\subsubsection{Without current relaxation}
Figure \ref{fig:currents} illustrates the evolution of the total current and current density for these scenarios without accounting for current relaxation. 
The high-Ne case, shown in Fig.~\ref{fig:currents}a and d, produces a runaway current density that is strongly localized in the core (mind the logarithmic scale of the colour-bar for $j_{\rm RE}$). Although the outer third of the plasma column is scraped off (represented by the black region in the upper right part of Fig.~\ref{fig:currents}d), this has practically no effect on the total runaway current evolution. 
This can be explained by the scrape-off process being, to some degree, self-reinforcing: The appearance of a RE plateau slows or stops the current decay, as well as the poloidal flux variation inside the plasma that would favour scrape-off. 
Thus, if a large RE population is generated sufficiently fast, scrape-off can not be expected to reach the core and mitigate a formed RE beam. 
 This parameter point is representative of the low-D--high-Ne region, where scrape-off remains ineffective without current relaxation, consistent with Figs.~\ref{fig:ire}a and c.

The low-D--low-Ne baseline case, shown in Fig.~\ref{fig:currents}b and e, exhibits a slower CQ due to lower resistivity. Without scrape-off the total current increases significantly after reaching a minimum at around $62\,\rm ms$, due to the longer ohmic time-scales allowing magnetic flux to diffuse across the resistive wall. (Most of this increase is outside the plotted time range.) This diffusion would in reality drive the development of massive RE currents during the nominal RE ``plateau'' phase, but, since the RE current continues increasing in this simulation (outside of the plotted domain), it does not form a true plateau. However with scrape-off, these longer timescales also provide more time for scrape-off to occur. The effect of scrape-off thus counteracts the RE generation driven by electric field diffusion, limiting the growth of the RE current.
Although $I_{\rm RE}$ still reaches almost $4\,\rm MA$, it does not grow significantly during this phase. By maintaining a constant magnetic flux at the last closed flux surface, the scrape-off process effectively neutralizes the impact of magnetic flux diffusion across the conducting wall. 

The high-D--low-Ne case, shown in Fig.~\ref{fig:currents}c and f, proves the most effective for RE suppression. The elevated deuterium content sustains a high effective critical electric field; thus even without scrape-off $I_{\rm RE}$ decays rapidly due to the increased collisional drag and peaks below $3\,\rm MA$. In this scenario, the initial significant RE population emerges closer to the edge. Here, localized cooling facilitates deuterium recombination at post-TQ temperatures, rendering scrape-off highly efficient at depleting the RE current. When scrape-off is enabled, the entire plasma column is exhausted in under $25\,\rm ms$. This rapid elimination contrasts sharply with the other evaluated cases, where the scrape-off process decelerates significantly once the RE current begins to dominate the total plasma current.   

\begin{figure}
    \begin{center}
        \includegraphics[width=11.88cm]{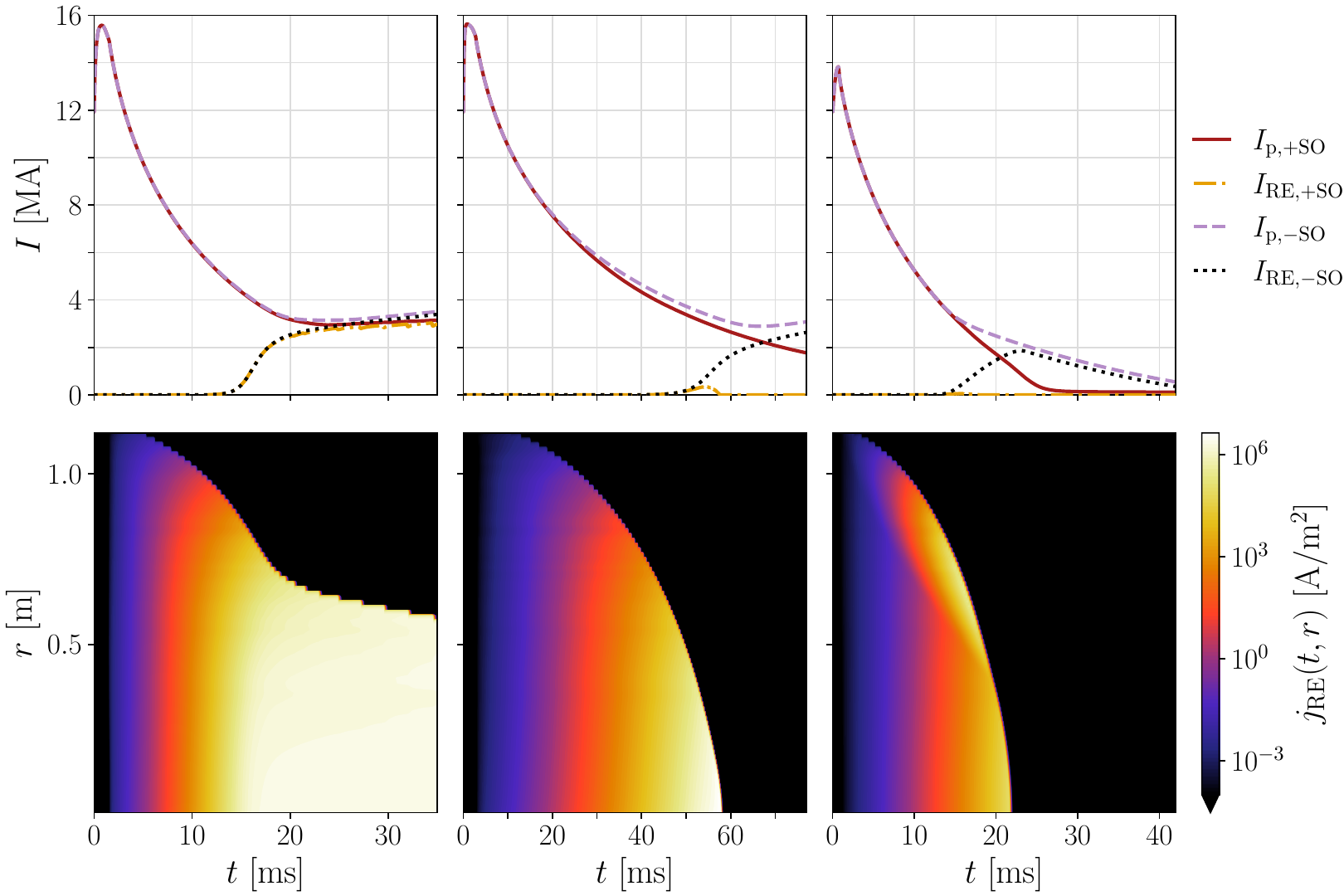}
        \put(-309,213){$\blacktriangle$} 
        \put(-215,213){\scalebox{0.8}{$\blacksquare$}}
        \put(-123,212){\scalebox{1.6}{$\bullet$}}
        \put(-243,212){(a)}
        \put(-150,212){(b)}
        \put(-56,212){(c)}
        \put(-243,106){\textcolor{white}{(d)}}
        \put(-150,106){\textcolor{white}{(e)}}
        \put(-56,106){\textcolor{white}{(f)}}
    \end{center}
    \caption{(a)-(c) Time evolutions of the total plasma currents and the runaway electron currents \textbf{with current density relaxation}, for the assimilated quantity combinations represented by the markers in Fig.~\ref{fig:ire}.  Solid and dashed curves are total plasma current, dash-dotted and dotted are runaway currents. Dashed and dotted are without, solid and dash-dotted are with scrape-off.   (a) low-D--high-Ne ($\blacktriangle$); (b) low-D--low-Ne baseline (\scalebox{0.8}{$\blacksquare$}); and (c) high-D--low-Ne (\raisebox{-0.3ex}{\scalebox{1.6}{$\bullet$}}) scenario. 
    (d)-(f) Spatio-temporal evolution of the runaway current density with scrape-off, for the respective assimilated densities of panels (a)-(c). }
\label{fig:currents2}    
    \end{figure}
\subsubsection{With current relaxation}
Next, we consider the same cases with the current relaxation accounted for, shown in Fig.~\ref{fig:currents2}, corresponding to Fig.~\ref{fig:ire}b and d. The first difference to observe is the initial $I_{\rm p}$-spike,  driven by a rapid current density redistribution (occurring faster than the resistive timescale) at a fixed magnetic helicity. This spike is approximately {$\SI{30}{\percent}$} for the low-D injection cases and {$\SI{15}{\percent}$} for the high-D case. Recalling that current relaxation is only kept active during the TQ, the reduced spike in the high-D  scenario is partly due to the shorter TQ interrupting the relaxation process. In addition, the MMI increases the plasma mass density, which decreases the Alfv\'en speed and extends the natural timescale of the current relaxation.

The low-D--high-Ne case remains qualitatively similar to the case without current relaxation, see Fig.~\ref{fig:currents2}a. However, now scrape-off provides a slight improvement by broadening the radial distribution of the runaway current, making it less centrally peaked.
The low-D--low-Ne baseline case exhibits a much more significant improvement. Although the RE current density remains core localized, the flattened current profile reduces the induced electric field at the end of the TQ, which weakens the initial drive for RE growth. This delay allows the entire plasma column to be scraped off before a massive core RE current can develop, resulting in a dramatic reduction in $I_{\rm RE}$ from more than {$\SI{4}{MA}$} (without scrape-off; the exact value slowly increases with time) to just $\SI{349}{kA}$ (with scrape-off). 
In the high-D case (Fig.~\ref{fig:currents2}f), the  RE current density remains near the plasma edge, enabling its efficient removal via scrape-off. Consequently, the maximum RE current drops even lower than in the case without current density relaxation. Furthermore, comparing Figs.~\ref{fig:ire}c and d reveals that the extreme high $n_{\text{D}}$ regime, where deuterium recombination previously aggravated RE generation, is pushed to even higher values of $n_{\text{D}}$. This favourable shift occurs because current flattening locally increases the edge ohmic current density, maintaining a higher post-TQ edge temperature. 
As a result, the synergistic combination of current relaxation and scrape-off successfully suppresses  $I_{\rm RE}$ to negligible levels in this regime.

If deuterium assimilation is constrained near the assumed limit of $n_{\rm D} = 1.4\times 10^{21}\,\rm{m}^{-3}$, maximizing deuterium assimilation becomes the optimal strategy to prevent an incomplete TQ. Similarly, the baseline neon concentration of $n_{\rm Ne} = 2.26 \times 10^{18} \, \si{m^{-3}}$ represents a suitable operating point; it is sufficiently high to avoid incomplete TQs while enabling current density relaxation and scrape-off to synergistically reduce $I_{\rm RE}$ to manageable levels. Increasing $n_{\rm D}$ is highly advantageous;  in such scenarios, holding $n_{\rm Ne}$ constant provides additional margin below the  $I_{\rm RE}$ reference value, whereas slightly increasing $n_{\rm Ne}$ along the $0.15\,\rm MA$ contour in Fig.~\ref{fig:ire}d lowers the conducted heat fraction while maintaining controlled RE currents.  With this approximate delineation of the interesting assimilation parameter space, simulations incorporating realistic injection, assimilation, and current profile evolution dynamics are required to find optimal operating scenarios, given the sensitivity of both $I_{\rm RE}$
  and the incomplete TQ boundary to transient plasma behaviour. 

\subsection{Avalanche gain factor}
\begin{figure}
\begin{center}
\includegraphics[width=11.88cm]{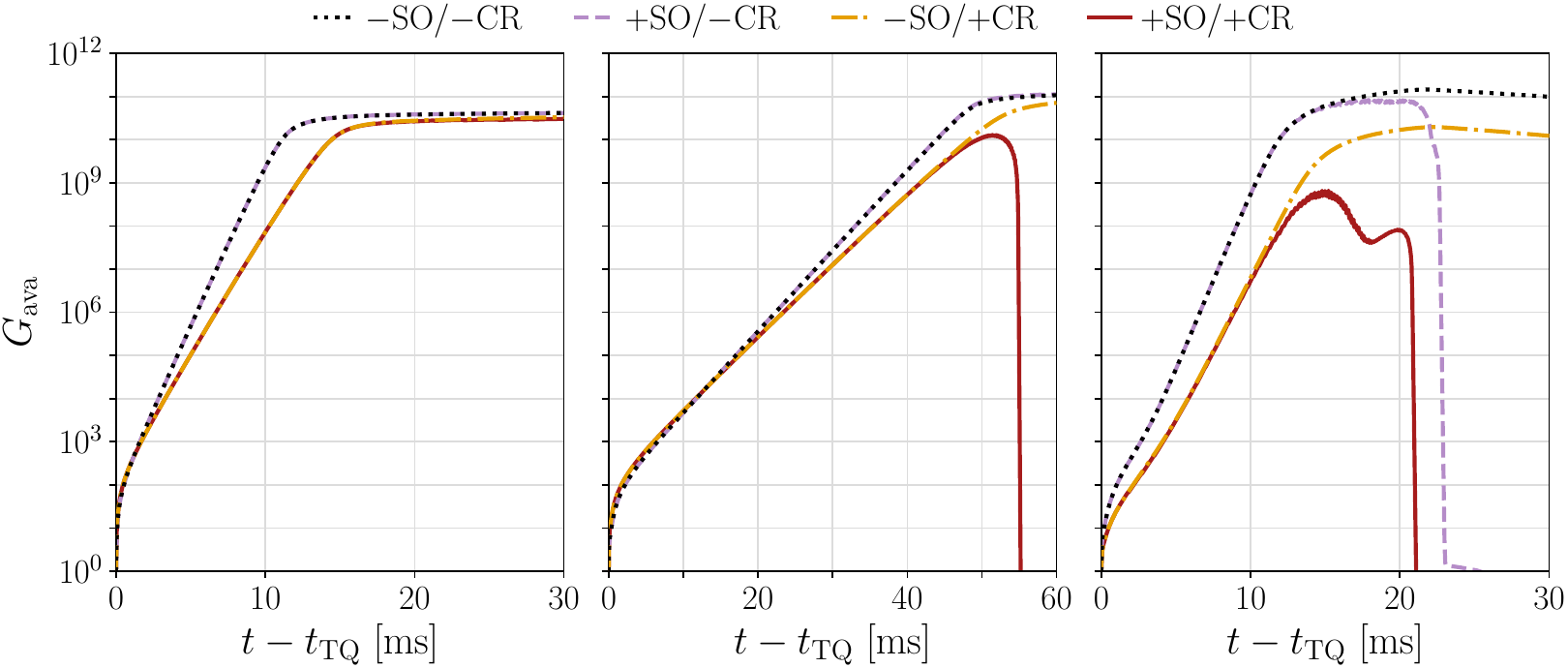}
\put(-309,124){$\blacktriangle$} 
\put(-202,124){\scalebox{0.8}{$\blacksquare$}}
\put(-97,123){\scalebox{1.6}{$\bullet$}}
\put(-310,25){(a)}
\put(-203,25){(b)}
\put(-98,25){(c)}
\end{center}
  \caption{ Time evolution of the avalanche gain $G_{\rm ava}$ for the (a) low-D--high-Ne ($\blacktriangle$); (b) low-D--low-Ne baseline (\scalebox{0.8}{$\blacksquare$}); and (c) high-D--low-Ne (\raisebox{-0.3ex}{\scalebox{1.6}{$\bullet$}}) scenario. 
  Dotted and dash-dotted curves are without, dashed and solid curves are with scrape-off; dotted and dashed curves are without, dash-dotted and solid are with current relaxation. 
 }
\label{fig:Gava}
\end{figure}

It is interesting to consider the avalanche gain factors, $G_{\rm ava}$, defined as the total runaway current normalized to the current at the end of the TQ, shown in Fig.~\ref{fig:Gava}.  As we have already established, scrape-off has practically no effect in the low-D--high-Ne case of Fig.~\ref{fig:Gava}a, due to the core localized RE population. However, now it is more clearly seen that the avalanche gain is reduced when current relaxation is accounted for, which is consistent with the reduced ohmic current in the core at the end of the TQ, and corresponding weaker electric field induction.

In the low-D--low-Ne baseline case of Fig.~\ref{fig:Gava}b, we find a weaker effect of the current relaxation and the scrape-off on the avalanche rate individually. 
Both effects have comparable slight reductions of the final RE current, due to the still core localized but somewhat wider current density profile being weakly impacted by either of these effects. 
However their combined effect not only has the reduced rate caused by the current flattening, but the maximum value is also strongly affected by a complete scrape-off of the plasma around the time the no-scrape-off case reaches its plateau.  

The effect of current relaxation appears earlier and remains more significant in the high-D--low-Ne case, than in the other two scenarios, translating to a significant reduction of the final RE current from current relaxation only. This early reduction, and corresponding delay, of the avalanche gain does not in itself explain why maximum RE current becomes smaller with current relaxation (without scrape-off), since the $G_{\rm ava}$ ratios are comparable to those seen in Fig.~\ref{fig:Gava}a. Here an additional difference is that the RE current maximum is reached at comparable times. Complete scrape-off is achieved independently of the current relaxation, while scrape-off is more efficient when it acts on a plasma with fewer runaways to begin with, since the scrape-off process is, as previously discussed, somewhat self-reinforcing.  
In the case with both scrape-off and current relaxation we also observe a double-peak in the evolution of $G_{\rm ava}$. This is because in this scenario there are two distinct RE populations, one core localized and one closer to the edge. The latter gets scraped off first, while the smaller, core localized population keeps growing a bit longer, but since it has no time to reach macroscopic values and back-react on the scrape-off process, it also gets scraped off eventually.   

\subsection{Scraped-off runaway current}
Finally, we consider the evolution of the scraped-off runaway electron current for the three representative scenarios. Accounting for scraped-off runaways is important because their energy is ultimately deposited in the wall. Even at relatively low energies, they therefore contribute to the wall heat load, while sufficiently energetic runaways may penetrate the surface and cause additional subsurface damage. Consequently, considering only the runaway current remaining within the plasma may not fully capture the potential impact on plasma-facing components. The resulting wall loads depend both on the time available for the runaways to gain energy before being scraped off and on the rate at which they are lost. Although our fluid treatment of the runaway population does not provide information on the energy spectrum of the lost runaways, it can nevertheless be used to obtain conservative estimates of their contribution to the wall heat load.

\begin{figure}
    \begin{center}  
    \includegraphics[width=11.88cm]{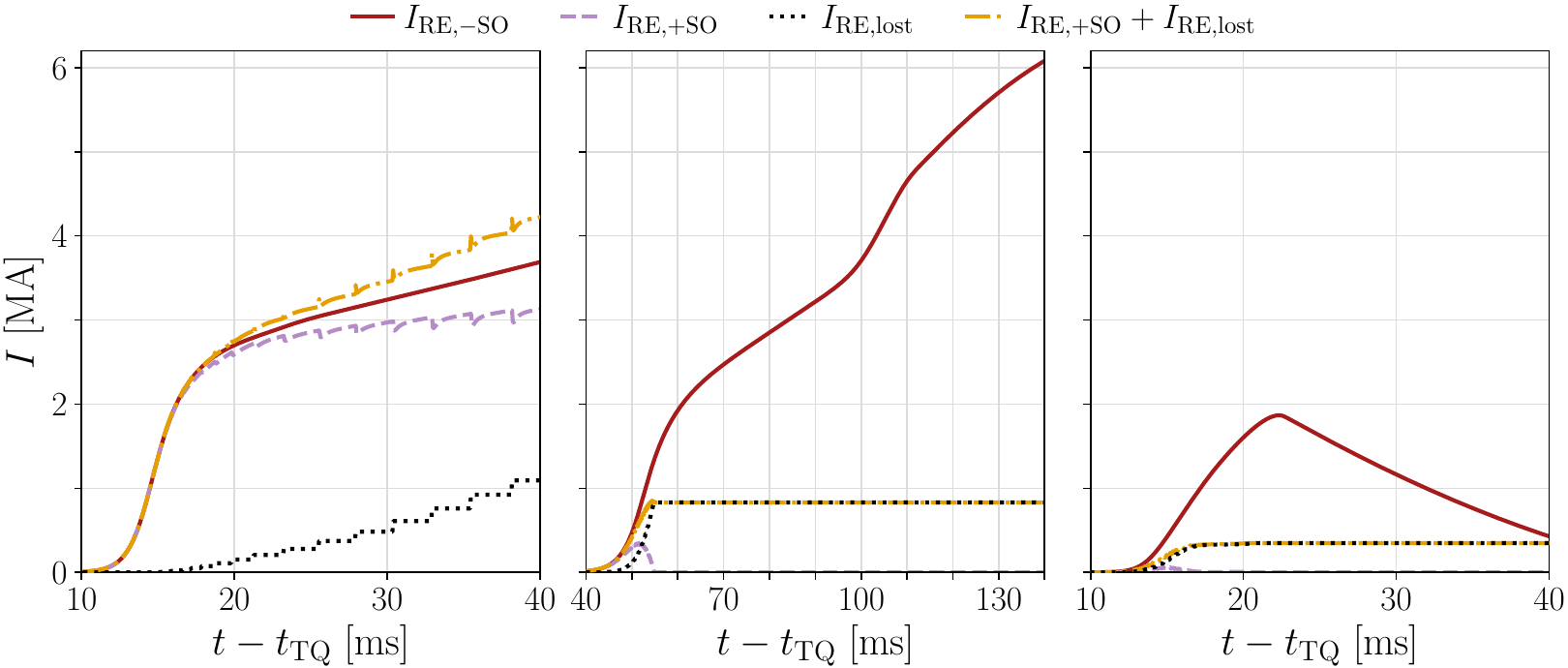}
\put(-317,124){$\blacktriangle$} 
\put(-208,124){\scalebox{0.8}{$\blacksquare$}}
\put(-100,123){\scalebox{1.6}{$\bullet$}}
    \put(-307,124){(a)}
    \put(-198,124){(b)}
    \put(-90,124){(c)}
    \end{center}   
\caption{Evolution of the scraped-off ($I_\text{RE,lost}$) and confined  ($I_\text{RE,+SO}$) runaway electron currents for the low-D-high-Ne (a), baseline low-D-low-Ne (b), and high-D-low-Ne (c) scenarios, illustrating the varying efficacy of the scrape-off mechanism, compared to the total runaway current generated without scrape-off ($I_\text{RE,-SO}$), as well as the sum of the scraped off and confined runaway currents ($I_\text{RE,+SO}+I_\text{RE,lost}$). Here, current relaxation is accounted for. }
\label{fig:scrapedoffRE}    
\end{figure}

Figure \ref{fig:scrapedoffRE} illustrates the varying impact of the scrape-off mechanism for different assimilated densities, by comparing the runaway current without scrape-off ($I_\text{RE,-SO}$, solid), the scraped-off runaway current ($I_\text{RE,lost}$, dotted), and the runaway current with scrape-off ($I_{\text{RE,+SO}}$, dashed). In the low-D--high-Ne scenario, Fig.~\ref{fig:scrapedoffRE}a, the combined total of the confined and scraped-off runaway currents follows, but is slightly higher than the runaway current generated in the absence of scrape-off, such that while the confined runaway current reaches approximately $3\,\rm MA$, the total of the confined and scraped-off runaways (dash-dotted) reaches above $4\,\rm MA$.  On the other hand, in the baseline low-D--low-Ne scenario, Fig.~\ref{fig:scrapedoffRE}b, the scrape-off mechanism dramatically reduces the total runaway generation. While the runaway current without scrape-off grows continuously beyond $4\,\rm MA$, accounting for scrape-off limits the total generated runaways to a plateau near $0.8\,\rm MA$. Finally, in the high-D--low-Ne scenario, Fig.~\ref{fig:scrapedoffRE}c, where high initial deuterium content limits the runaway growth to approximately $2\,\rm MA$ without scrape-off, applying the scrape-off model suppresses the confined runaway current almost entirely and restricts the total scraped-off current to $0.3\,\rm MA$.

In the cases shown in Fig.~\ref{fig:scrapedoffRE}b and c, the
cumulative lost runaway current, $I_{\text{RE,lost}}$, exceeds
the maximum confined runaway current. This is possible because
$I_{\text{RE,lost}}$ accumulates losses over time, whereas
$I_{\text{RE,+SO}}$ represents the instantaneous confined
runaway current. For example, if runaway generation in the confined
region balances scrape-off losses, the confined runaway current
remains constant while $I_{\text{RE,lost}}$ continues to increase
and can exceed it.
In calculating $I_{\text{RE,lost}}$, we accumulate losses only
while a finite confined region remains. Accordingly, in
Fig.~\ref{fig:scrapedoffRE}b and c, $I_{\text{RE,lost}}$ stops
increasing once the plasma has been completely scraped off and
$I_{\text{RE,+SO}}$ vanishes.
A plateau in $I_{\text{RE,lost}}$ therefore indicates that no
further losses are being accumulated, rather than the persistence
of a runaway beam. Its time derivative,
$\mathrm{d}I_{\text{RE,lost}}/\mathrm{d}t$, measures the scrape-off
loss rate in current-equivalent units and characterizes the temporal
concentration of these losses. This rate is relevant to assessing
wall loads, together with the energy distribution of the lost
electrons and their deposition footprint. A comparison with the
wall loads from a subsequent impact of the remaining confined beam
would additionally require its loss timescale and impact
characteristics, which are not modelled here.

\begin{figure}
    \begin{center}  
    \includegraphics[width=11.88cm]{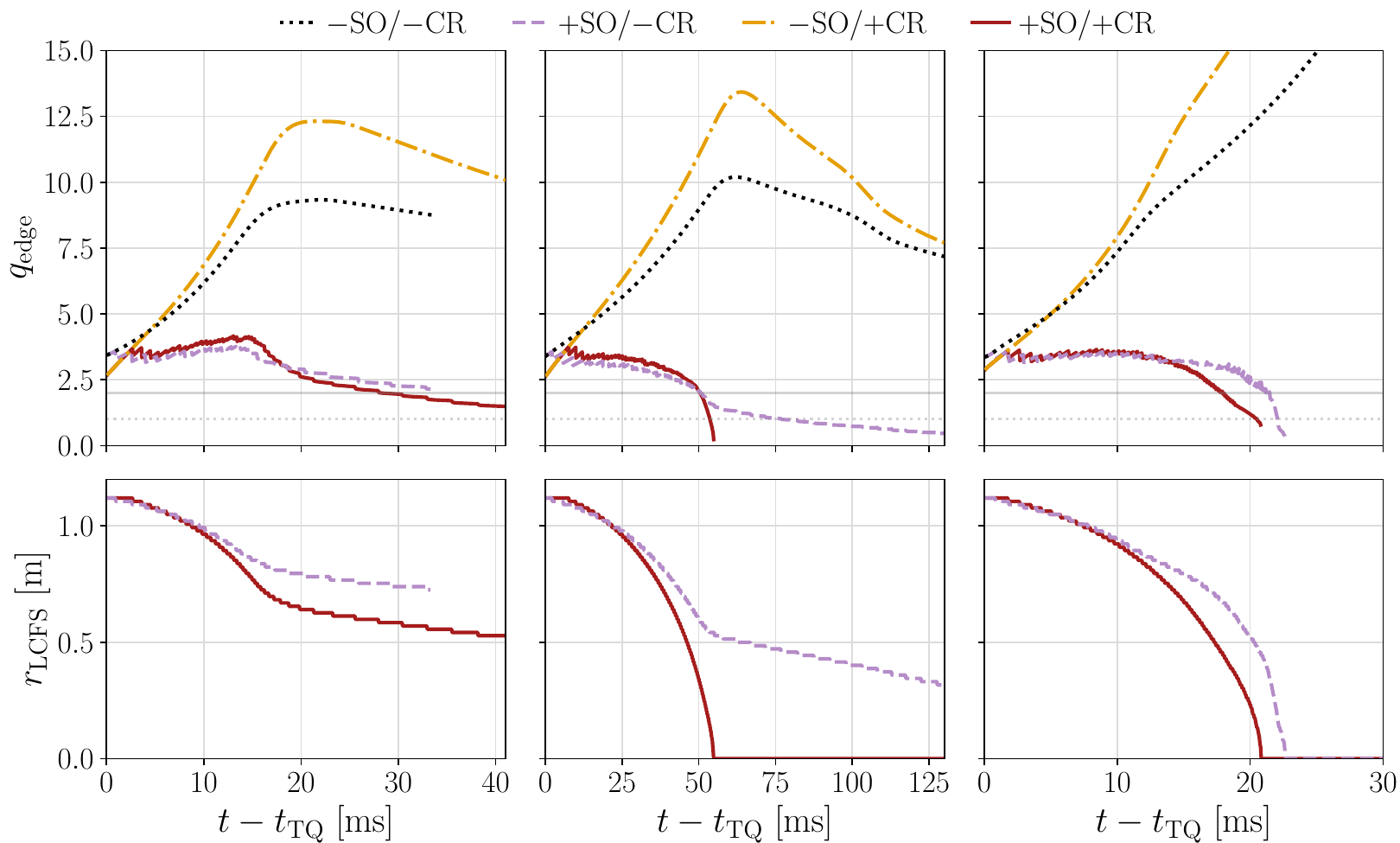}
    \put(-310,182){$\blacktriangle$}
    \put(-203,182){\scalebox{0.8}{$\blacksquare$}}
    \put(-98,182){\raisebox{-0.3ex}{\scalebox{1.6}{$\bullet$}}}
    \put(-230,182){(a)}
    \put(-124,182){(b)}
    \put(-17,182){(c)}
    \put(-230,79){(d)}
    \put(-124,79){(e)}
    \put(-17,79){(f)}    
    \end{center}
\caption{Time evolution of (a)--(c) the edge safety factor $q_\mathrm{edge}$, evaluated at the instantaneous LCFS, and (d)--(f) its minor radius $r_\mathrm{LCFS}$, for the (a,d) low-D--high-Ne ($\blacktriangle$), (b,e) low-D--low-Ne baseline (\scalebox{0.8}{$\blacksquare$}), and (c,f) high-D--low-Ne (\raisebox{-0.3ex}{\scalebox{1.6}{$\bullet$}}) scenarios. Dotted and dash-dotted curves correspond to simulations without scrape-off, while dashed and solid curves include scrape-off. Dotted and dashed curves omit current relaxation, while dash-dotted and solid curves include it. In panels (a)--(c), the gray solid and dotted horizontal lines mark $q_\mathrm{edge}=2$ and $1$, respectively. Without scrape-off, $r_\mathrm{LCFS}$ remains fixed at the initial minor radius, so only the $\text{+SO}$ cases are shown in panels (d)--(f).
}
\label{fig:qedge}    
\end{figure}

The evolution of the edge safety factor, evaluated at the instantaneous LCFS, is shown in Figure~\ref{fig:qedge} together with the radius of the LCFS. Without scrape-off, the plasma boundary is fixed and $q_\mathrm{edge}$ increases substantially during the current quench as the total plasma current decays. With scrape-off, by contrast, the contraction of the confined cross-section keeps $q_\mathrm{edge}$ low; after an initial phase at $q_\mathrm{edge}\approx 3$--$4$, it decreases as the LCFS moves inward. In the low-D--high-Ne case ($\blacktriangle$), the boundary contraction slows down significantly once the runaway current dominates, leaving a confined region of radius $r_\mathrm{LCFS}\approx 0.55$--$0.75\,\mathrm{m}$, while $q_\mathrm{edge}$ reaches approximately $1.5$ (+CR) and $2$ (-CR) during these simulations. In the baseline (\scalebox{0.8}{$\blacksquare$}) and high-D (\raisebox{-0.3ex}{\scalebox{1.6}{$\bullet$}}) cases, $q_\mathrm{edge}$ drops below $2$ and eventually below unity. The plasma is eventually scraped off completely in these cases, except in the baseline $\text{+SO/-CR}$ simulation.  This suggests that additional current-driven MHD instabilities could affect the subsequent evolution and runaway losses. The onset and consequences of such instabilities depend on the current profile and plasma--wall coupling and cannot be determined from $q_\mathrm{edge}$ alone. These effects are not resolved by the present reduced model.

\section{Conclusions}
\label{sec:concl}

 By incorporating both a hyperresistive current relaxation model and a flux-conservative scrape-off model in the DREAM fluid-kinetic framework, MMI scenarios were evaluated using varying densities of deuterium and neon in ARC. The simulations demonstrated that operating under the conservative assumption of a fixed magnetic flux-surface geometry leads to an overestimation of the final runaway electron current. 
 
The inclusion of scrape-off effects alone yields modest reductions in runaway currents, particularly in parameter regimes with high deuterium and low neon densities where the critical electric field remains high. However, combining scrape-off with hyperresistive current flattening creates a highly beneficial synergistic effect. Hyperresistive current relaxation broadens the current channel outward, shifting efficient runaway electron generation closer to the plasma edge. This allows the dynamically contracting boundary, dictated by the vertical displacement, to more effectively scrape off the electron beams. This interplay significantly expands the operational window for successful disruption mitigation toward lower assimilated densities and reduces the confined runaway current to more manageable levels.

Despite the improved runaway suppression, simultaneously meeting all three reference values introduced in Section~\ref{sec:intro} remains challenging. No region of the explored parameter space satisfies all three simultaneously:  a maximum runaway
electron current below $150\,\rm kA$, a conducted heat fraction below $10\%$,
and a current quench duration between $11.3\,\rm ms$ and $142\,\rm ms$.
Relaxing the runaway current reference value to approximately $200\,\rm kA$
already allows the regions meeting the current and heat-loss reference values
to overlap.

Note that these conclusions regarding the conducted heat fraction should be
regarded as conservative. As noted in Section~2, the electron-thermal-speed
scaling of the parallel heat flux used here likely overestimates conducted
losses \citep{Zhang2023}, and \citet{ARC2026} found that the
sheath-limited heat flux boundary condition of \citet{Kiramov2016}
substantially reduces the conducted heat fraction while leaving the runaway
electron dynamics largely unchanged. A more restrictive treatment of parallel heat losses could therefore extend the region satisfying the $\eta_{\rm cond}$ reference value towards lower assimilated densities. Moreover, $\eta_{\rm cond}$ is only a proxy for the relevant engineering
criterion, avoidance of melting of plasma-facing components, which also
depends on the spatial and temporal distribution of the deposited heat.

The present study uses a fixed representative value of $\delta B/B$ to isolate the combined effects of current flattening and scrape-off. Although current relaxation is in the saturated regime at this value (Appendix~\ref{sec:AppendixSpatialVariation}), the total runaway current and conducted heat loss remain sensitive to TQ transport \citep{Pusztai_2023}. Changes in $\delta B/B$ could therefore shift the quantitative mitigation boundaries through their effects on plasma cooling and runaway seeding, even though scrape-off occurs predominantly after the TQ. This sensitivity is not explored here.

Finally, we note that achieving a viable disruption mitigation strategy will require more advanced simulations that incorporate detailed modelling of MMI dynamics, impurity assimilation, and precise current density profile evolution. Furthermore, the scraped-off runaway current can substantially exceed
the maximum confined runaway current. Assessing the resulting wall
loads will require modelling the energy distribution of the lost
electrons and their spatial and temporal deposition on the wall.

\section*{Acknowledgements}
The authors are grateful  to Mathias Hoppe, Max Larsson, Oskar Vallhagen, Lorenzo Votta and Björn Zaar for fruitful discussions.
The computations were enabled by resources provided by the National Academic Infrastructure for Supercomputing in Sweden (NAISS), partially funded by the Swedish Research Council through grant agreement no.\ 2022-06725. The work has been carried out within the
framework of the EUROfusion Consortium, funded by the European Union via the
Euratom Research and Training Programme (Grant Agreement No 101052200 —
EUROfusion). Views and opinions expressed are however those of the authors
only and do not necessarily reflect those of the European Union or the European
Commission. Neither the European Union nor the European Commission can be
held responsible for them.
\section*{Declaration of Interests}
The authors report no conflict of interest.
\section*{Funding}
This work was supported by the Swedish Research Council (Dnr.~2022-02862), the Swedish Energy Agency, and by the Knut and Alice
Wallenberg foundation (Dnrs.~2022.0087 and 2023.0249).  The authors affiliated with Commonwealth Fusion Systems (CFS) and MIT are funded by CFS.

\appendix

\section{Effect of radially varying hyperresistivity}
\label{sec:AppendixSpatialVariation}

To assess the sensitivity of the results to a possible radial variation of $\Lambda$, we use the profile
\begin{equation}\label{eq:lambda_profile}
\Lambda(r) = \Lambda_0 \frac{p+1}{2^{p+1}-1}\bigg[1 + \bigg(\frac{r}{a}\bigg)^2\bigg]^p
\end{equation}
which fulfils the constraint that the average $\int_0^a \mathrm{d}r \, r \Lambda(r) \big/ \int_0^a \mathrm{d}r \, r$ is constant and given by the estimate Eq.~(\ref{vottalambda}), here denoted by $\Lambda_0$.
Profiles with $p=0$ are radially constant, whereas profiles with $p<0$ ($p>0$) are biased towards the core (edge). The latter may be used to emulate the effect of an outward-increasing magnetic perturbation level due to externally applied magnetic perturbations.   For reference, recalling our choice $\delta B/B=0.3\%$, taking $\psi_{\rm t}$ at the plasma edge as a representative value, and calculating $V_{\rm A}$ for the actual plasma composition,   the low-D--low-Ne baseline case
(\scalebox{0.8}{$\blacksquare$}) corresponds to $\Lambda \approx \SI{0.91}{Wb^3/C}$.

\begin{figure}
\centering
\includegraphics[width=11.66cm]{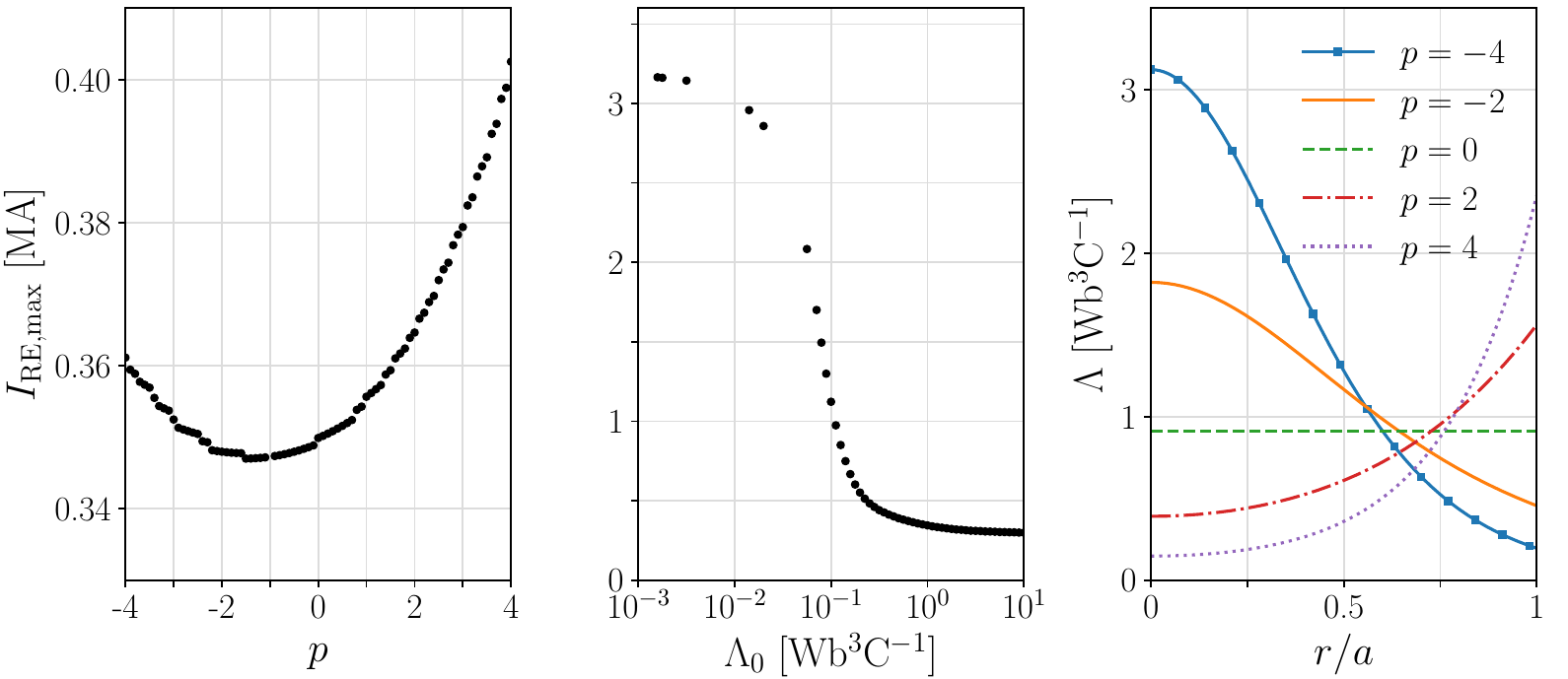}
\put(-303,132){(a)}
\put(-194,132){(b)}
\put(-86,132){(c)}
\caption{Sensitivity of the maximum runaway current to (a) radial variations in $\Lambda$, parametrized by $p$, at fixed $\Lambda_0 \approx 0.91\,\rm Wb^3/C$, and (b) the magnitude of a radially uniform $\Lambda$. Both scans correspond to the low-D--low-Ne baseline case (\scalebox{0.8}{$\blacksquare$}). Note the different vertical-axis ranges. Panel (c) illustrates the radial profiles of $\Lambda$ corresponding to different values of $p$.  }
\label{fig}
\end{figure}

Figure~\ref{fig}a shows the dependence of the maximum runaway current
on $p$ for the baseline case. The details of how such a representative value the maximum runaway current is calculated are given in Appendix A of \citep{ekmark2024}. The dependence is approximately
parabolic, with a minimum for a $\Lambda$ profile slightly weighted
towards the core. Despite the wide range of $p$ considered,
$I_{\rm RE,max}$ varies by only about $15\%$. This weak sensitivity
arises because the adopted value of $\Lambda_0$ is sufficient to
almost completely flatten the current density profile, even when
$\Lambda$ varies radially.

Figure~\ref{fig}b illustrates this saturation by varying the magnitude
of a radially uniform $\Lambda$. Most of the reduction in
$I_{\rm RE,max}$ occurs as $\Lambda$ increases through the range
$0.01$--$0.1\,\rm Wb^3/C$, with progressively weaker sensitivity
at larger values. The baseline value,
$\Lambda \approx 0.91\,\rm Wb^3/C$, lies well within this regime
of weak sensitivity.

\bibliographystyle{jpp}
\bibliography{refs} 

\end{document}